\documentclass{ws-procs975x65}

\newcommand{\beq}{\begin{equation}}
\newcommand{\eeq}{\end{equation}}
\newcommand{\beqs}{\begin{eqnarray}}
\newcommand{\eeqs}{\end{eqnarray}}
\newcommand{\pbp}[0]{\ensuremath{\langle \overline{\psi} \psi \rangle}}
\newcommand{\mres}{m_{\rm res}}
\newcommand{\chidof}{\ensuremath{\mbox{$\chi^2/\text{d.o.f.}$}}}
\newcommand{\MSbar}[0]{\overline{\text{MS}}}

\begin{document}

\title{From Lattice Strong Dynamics to Phenomenology}

\author{George T.~Fleming$^{a}$ and Ethan T.~Neil$^{b,c,d}$\footnote{Speaker.  E-mail:ethan.neil@colorado.edu}\\
\vspace{2mm}(for the Lattice Strong Dynamics (LSD) Collaboration)}

\address{$^a$ Department of Physics, Sloane Laboratory, Yale University, \\
New Haven, CT 06520, USA \\
$^b$Theoretical Physics Group, Fermi National Accelerator Laboratory, \\
Batavia, IL 60510, USA \\
$^c$Department of Physics, University of Colorado, \\
Boulder, CO 80309, USA \\
$^d$RIKEN-BNL Research Center, Brookhaven National Laboratory, \\
Upton, NY 11973, USA}

\begin{abstract}
We present updated results on the chiral properties of SU$(3)$ gauge theories with $N_f = 2$ and $6$ massless Dirac fermions in the fundamental representation.  Our focus is on the ratio $\pbp / F^3$, where $\pbp$ is the chiral condensate and $F$ is the pseudo-Nambu-Goldstone-boson decay constant.  This ratio is of interest in the context of fermion mass generation within composite Higgs theories.  By re-expanding certain ratios using next-to-leading-order chiral perturbation theory, we confirm our previous result of significant enhancement of this ratio at $N_f = 6$ over $N_f = 2$.
\end{abstract}

\keywords{SCGT12; chiral condensate; many-fermion theories; lattice simulation.}

\bodymatter

\section{Introduction}

Composite Higgs theories, in which a new strongly-coupled sector is responsible for the dynamical breaking of electroweak symmetry, are an attractive possibility for physics beyond the Standard Model.  These extensions of the Standard Model have a number of attractive theoretical features, in particular resolving the naturalness problem of the fundamental Higgs boson; as a composite bound state, the Higgs mass is generated from the underlying strong dynamics, and is not subject to quadratically divergent quantum corrections.  

Lattice simulation plays a crucial role in the study of composite Higgs theories, by allowing rigorous non-perturbative calculations in strongly-coupled gauge theories other than QCD.  The properties of these theories can vary greatly depending on the matter content; in particular, with a large number of fermions, the long-distance confinement phase of the theory is lost, giving way to an infrared-conformal phase in which chiral symmetry remains unbroken \cite{Caswell:1974gg,Banks:1981nn}.  In recent years, several lattice groups have turned their attention to the study of these gauge theories beyond QCD \cite{Neil:2012cb,Giedt:2012it}.

In addition to exploring theoretical issues such as the nature and location of the transition from confining to infrared conformal, lattice simulations can be used to calculate phenomenologically relevant quantities for composite Higgs models from first principles.  One quantity of particular interest is the chiral condensate $\pbp$.  In the context of a composite Higgs model, the condensate plays the role of the Higgs vacuum expectation value, giving rise to fermion masses through four-fermion couplings $\bar{\psi} \psi \bar{f} f$, in a way analogous to the standard Higgs-Yukawa mechanism.  In general, to satisfy experimental constraints based on flavor-changing neutral currents, a large value of $\pbp$ relative to the symmetry-breaking scale $F$ is needed, a phenomenon which is conjectured to occur near the critical number of fermions for the infrared-conformal transition \cite{Holdom:1981rm,Yamawaki:1985zg,Appelquist:1986an}.

We present here updated results \cite{Appelquist:2009ka} for the quantity $\pbp / F^3$ calculated in SU$(3)$ gauge theory, with $N_f = 2$ and $6$ Dirac fermions in the fundamental representation.  Our main improvement is a new technique for chiral extrapolation, which allows a more precise determination of the ratio of $\pbp / F^3$ between the two theories.

\section{The chiral condensate}

To make the connection to composite Higgs theories, we are interested in studying the chiral limit $m \rightarrow 0$, in which the electroweak symmetry is not broken explicitly.  Direct extraction of the chiral condensate in the limit $m \rightarrow 0$ from lattice calculations is challenging due to the presence of a quadratic divergence in the lattice cutoff contained in the linear term \cite{Neil:2010sc}.

We can make use of additional indirect information on the condensate by working in chiral perturbation theory \cite{Gasser:1984gg}.  At leading order, the Gell-Mann-Oakes-Renner (GMOR) relation gives the chiral condensate in terms of the pseudo-Goldstone boson mass $M$ and decay constant $F$,
\beq
M_m^2 F_m^2 = 2m \pbp_m,
\eeq
where the subscript $m$ denotes evaluation at finite fermion mass $m$, to distinguish from the chiral-limit values of $F$ and $\pbp$.

At higher order, re-expanding in the parameter $z \equiv 2B/(4\pi F)^2 = 2\pbp / (4\pi)^2 F^4$, we may re-write the mass dependence of $M_m$ \cite{Gasser:1986vb, Appelquist:2009ka, Neil:2010sc}
\begin{align}
\frac{M_m^2}{2m} &= 8\pi^2 F^2 z \left\{ 1 + zm\left[ \alpha_M + \frac{1}{N_f} \log(zm)\right] \right\}, \\
F_m &= F \left\{ 1 + zm \left[ \alpha_F - \frac{N_f}{2} \log(zm)\right] \right\}, \\
\pbp_m &= 8\pi^2 F^4 z \left\{ 1 + zm\left[ \alpha_c - \frac{N_f^2 - 1}{N_f} \log(zm)\right] \right\}.
\end{align}
From these formulas it is easy to see that chiral perturbation theory is an expansion in $zm = M_P^2 / (4\pi F)^2 = M_P^2 / \Lambda_\chi^2$.

We are now interested in constructing ratios of observables which will reduce to $\pbp / F^3 = 8\pi^2 Fz$ in the chiral limit.  There are three such constructions:
\begin{align}
X_m^{(FM)} &= \frac{M_m^2}{2mF_m}, \\
X_m^{(CM)} &= \frac{(M_m^2/2m)^{3/2}}{\pbp_m^{1/2}}, \\
X_m^{(CF)} &= \frac{\pbp_m}{F_m^3}.
\end{align}
Expanding these ratios out using the NLO expressions above, we recover their mass dependence to $\mathcal{O}(zm)$:
\begin{align}
X_m^{(FM)} &= 8\pi^2 F z \left(1 + zm\left[  \alpha_M - \alpha_F + \frac{N_f^2 + 2}{2N_f} \log(zm)\right] \right) \\
X_m^{(CM)} &= 8\pi^2 F z \left(1 + zm \left[ \frac{3}{2} \alpha_M - \frac{1}{2} \alpha_C + \frac{N_f^2 + 2}{2N_f} \log(zm)\right] \right) \\
X_m^{(CF)} &= 8\pi^2 F z \left(1 + zm \left[ \alpha_C - 3\alpha_F + \frac{N_f^2 + 2}{2N_f} \log(zm) \right] \right)
\end{align}
By construction, all three ratios have the same intercept $8\pi^2 F z = \pbp / F^3$.  The coefficient of the chiral logarithm in each expansion is identical, but this is a coincidence, and does not persist for the leading terms at next-to-next-to-leading order (NNLO) \cite{Bijnens:2009qm}.

For comparison of results between the $N_f = 2$ and $N_f = 6$ theories, we construct the ``ratio of ratios"
\beq
R_{\tilde{m}}^{(XY)} \equiv \frac{ X_m^{(XY)}(N_f=6)}{ X_m^{(XY)} (N_f=2)},
\eeq
where $(XY)$ enumerates the three possible constructions $(FM), (CM), (CF)$.  The ratio is taken at constant bare mass $m_f$, but the renormalized masses differ slightly due to the variation of $\mres$; we thus extrapolate in the geometric mean mass, $\tilde{m} \equiv \sqrt{m_{N_f=2} m_{N_f=6}}$.  As shown in appendix A, expanding $R$ to NLO in $\tilde{m}$ yields the general functional form
\beq
R_{\tilde{m}}^{(XY)} = R^{(6)}(1 + \alpha_R^{(XY)} \tilde{m} + \beta_R^{(XY)} \tilde{m} \log \tilde{m} + ...),
\eeq
where the intercept $R^{(6)}$ is independent of the particular ratio $XY$ used in the construction.

Finally, we note that it is possible to construct additional ratios in terms of other quantities related to the PNGB.  For example, the $I=2$ PNGB scattering length $a_{PP}$ is proportional at leading order to $(M a_{PP}) \propto -M^2 / (16\pi F^2)$ \cite{Appelquist:2012sm}, so that the appropriate combination is
\beq \label{eq:rAFM}
X_m^{(AFM)} = \frac{8\pi F_m M_m a_{PP,m}}{m}.
\eeq
Other constructions are possible.  The statistical precision of the scattering length $a_{PP}$ is typically not comparable to the other observables considered here, so we will not include this ratio in our chiral extrapolation to determine $R^{(6)}$.  However, agreement between ratios such as this and the more precise set of ratios given above is a useful consistency check.

\section{Results}

Our lattice calculations are performed using $32^3 \times 64$ domain-wall lattices with the Iwasaki improved gauge action.  The gauge coupling is set to $\beta = 2.70$ at $N_f = 2$, and $\beta = 2.10$ at $N_f = 6$.  These $\beta$ values are tuned in order to match the confinement scale of the two theories in the chiral limit.  Further details of the simulation are given in a prior reference \cite{Appelquist:2009ka}.

For both $N_ f = 2$ and $N_f = 6$, results for six mass points are shown.  All chiral extrapolations are considered in terms of the total fermion mass $m \equiv m_f + m_{res}$.  We present results for the lightest point $m_f = 0.005$, but do not include them in the analysis due to the likely presence of uncontrolled finite-volume systematic effects.  Other simulation results based on these gauge configurations have been presented in previous references \cite{Appelquist:2010xv,Appelquist:2012sm,Appelquist:2013ms}.

\begin{figure}
\begin{center}
\includegraphics[width=0.8\textwidth]{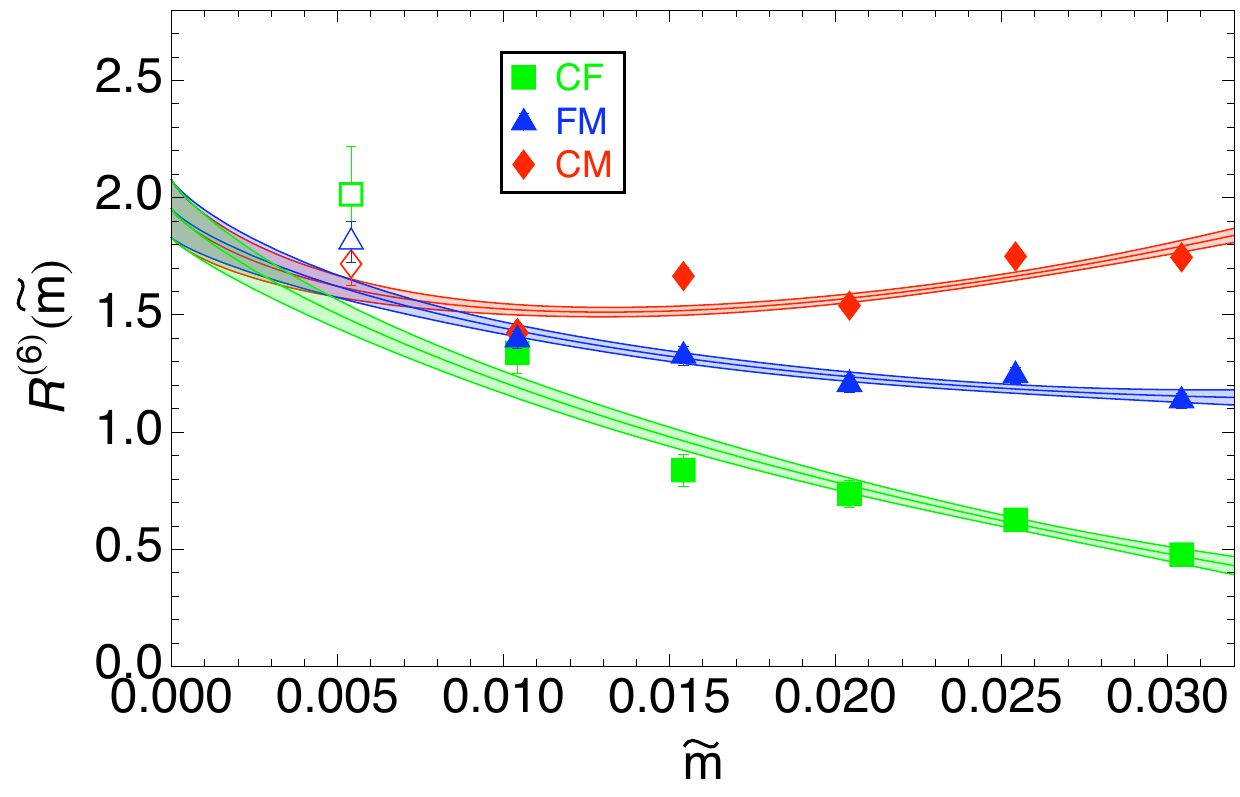}
\caption{Extrapolation of $R^{(6)}$ to the chiral limit.  ``CF", ``FM" and ``CM" label the three different combinations of observables which all extrapolate to $R^{(6)}$ in the chiral limit, as discussed in the text.  The best fit shown is a joint fit to all closed symbols, constrained to have a common intercept value, and includes $1\sigma$ error bands.  The open symbols at $m = 0.005$ are not included in the analysis, due to possible uncontrolled systematic errors. \label{fig:R2f6f}}
\end{center}
\end{figure}

Determination of pseudoscalar masses and decay constants, along with the chiral condensate, is discussed and data tables are presented in a previous work \cite{Appelquist:2009ka}.  Data and joint fit results for the three constructed ``ratios of ratios" are shown in Fig.~\ref{fig:R2f6f}.  The joint fit shown has $\chidof = 5.73$ with $8$ degrees of freedom; the relatively large value of $\chi^2$ is likely due to autocorrelation effects in the simulation data, as discussed previously \cite{Appelquist:2009ka}.  The intercept of the fit is
\beq
R^{(6)}(\Lambda) = 1.95(12).
\eeq
This quantity is unrenormalized.  Applying one-loop lattice perturbation theory to compute the renormalization factors for the chiral condensate \cite{Aoki:2002iq}, we find the correction factor \cite{Appelquist:2009ka} $Z_6 / Z_2 = 0.8527(97)$, so that
\beq
R^{(6)}_{\MSbar}(\Lambda) = 1.60(10).
\eeq
The fact that $R^{(6)} > 1$ indicates enhancement of the chiral condensate relative to the $N_f = 2$ case.  In particular, this is a substantial and statistically significant increase compared to the perturbative estimate \cite{Appelquist:2009ka}, $R^{(6)}_{\MSbar} \leq 1.15$.

\begin{figure}
\begin{center}
\includegraphics[width=0.8\textwidth]{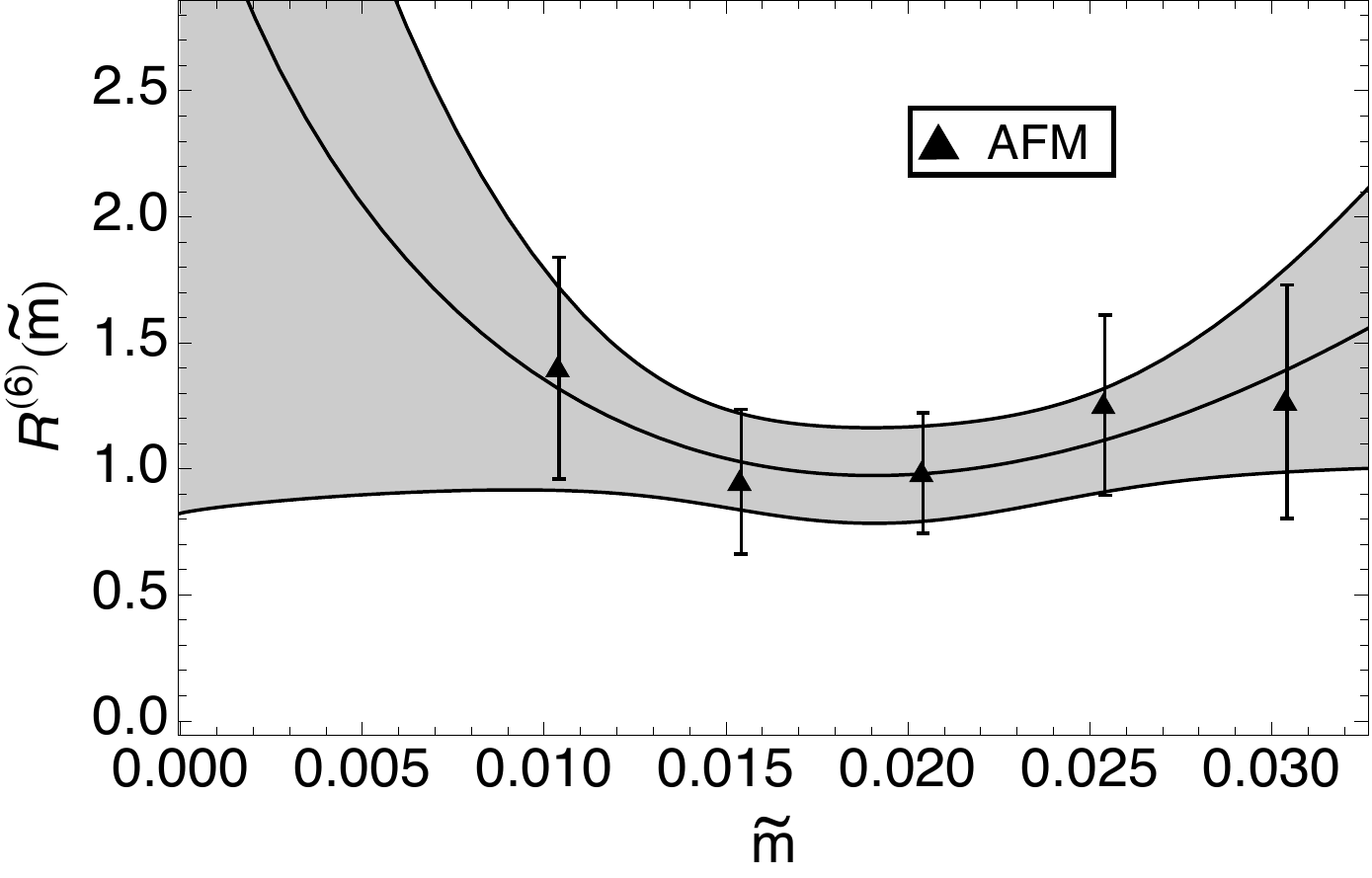}
\caption{Alternate determination of $R^{(6)}$ using the pseudoscalar scattering length $a_{PP}$, as given in eq.~\ref{eq:rAFM}.  A chiral extrapolation with $1\sigma$ error bands is also shown.  Although the precision of $R^{(6)}$ is much lower in this channel, consistency of the data points with fig.~\ref{fig:R2f6f} gives a useful cross-check. \label{fig:RAFM}}
\end{center}
\end{figure}

\clearpage

\begin{table}[t]
\begin{center}
\tbl{Simulation results for $N_f = 2$ and $6$.  The lightest points at $m_f = 0.005$ are excluded from our analysis as discussed in the text.}{\begin{tabular}{|cc|cccc|}
\hline
$N_f$&$m$&$M_m^2$&$F_m$&$\pbp_m$&$M_\rho$\\
\hline
\hline
2&\textit{0.005026}&\textit{0.01183(32)}&\textit{0.02468(55)}&\textit{0.0058293(90)}&\textit{0.2449(60)}\\
&0.010026&0.01917(18)&0.03080(39)&0.0109544(47)&0.2553(30)\\
&0.015026&0.02735(30)&0.03294(69)&0.0160535(52)&0.2607(32)\\
&0.020026&0.04060(45)&0.03586(64)&0.0211960(66)&0.2998(34)\\
&0.025026&0.04798(54)&0.03709(57)&0.0262204(48)&0.3100(28)\\
&0.030026&0.05947(54)&0.03795(52)&0.0312388(72)&0.3241(23)\\
\hline
6&\textit{0.005823}&\textit{0.02124(49)}&\textit{0.02113(52)}&\textit{0.007365(15)}&\textit{0.2472(41)}\\
&0.010823&0.02798(27)&0.02989(53)&0.013395(6)&0.2750(32)\\
&0.015823&0.04304(44)&0.03718(64)&0.019327(9)&0.3019(37)\\
&0.020823&0.05959(50)&0.04207(83)&0.025179(9)&0.3429(33)\\
&0.025823&0.07595(44)&0.04584(90)&0.030964(9)&0.3755(33)\\
&0.030823&0.09331(56)&0.05120(106)&0.036640(10)&0.4040(35)\\
\hline
\end{tabular}}
\end{center}
\label{table:data}
\end{table}

\section{Conclusion}

We have presented updated results for enhancement of the condensate ratio $\pbp / F^3$ in SU$(3)$ gauge theories with $N_f$ fundamental fermions, as $N_f$ is increased from $2$ to $6$.  This ratio is important in composite Higgs theories, as the size of the chiral condensate is related to the generation of fermion masses.  In general, a large value of the condensate is needed to satisfy phenomenological constraints; the ratio is conjectured to increase dramatically for many-fermion theories near (but below) the transition value $N_f^c$ to infrared-conformal behavior \cite{Holdom:1981rm,Yamawaki:1985zg,Appelquist:1986an}.

Our lattice results confirm a significant increase of the enhancement at $N_f = 6$; the ``ratio of ratios" comparing $N_f = 6$ to $N_f = 2$ is found to be $R^{(6)}_{\MSbar} = 1.60(10).$  This is large relative to a perturbative estimate, $R^{(6)}_{pert} \lesssim 1.15$.  Future work at larger values of $N_f$ closer to the critical value for transition for loss of confinement is in progress, and will provide further information on this interesting trend.

\section*{Acknowledgments}

The LSD collaboration would like to thank the LLNL Multiprogrammatic and institutional Computing program for Grand Challenge allocations and time on the LLNL BlueGene/L (uBGL) supercomputer.  We thank LLNL for funding from LDRD 10-ERD-033 and LDRD 13-ERD-023.  This work has been supported by the National Science Foundation under Grant No.~NSF PHY11-00905 (G.~F.).  Fermilab is supported by Fermi Research Alliance, LLC under Contract No.~DE-AC02-07CH11359 with the U.~S.~Department of Energy.

\clearpage

\appendix{Chiral expansion of ratios with mistuned fermion masses}

Comparison of results between two different theories generally requires careful tuning to match physical scales.  However, in our simulation results there is a slight mistuning of the fermion masses, due to the residual domain-wall contribution $\mres$.  We can correct for mistuning of the fermion mass by using chiral perturbation theory.  Suppose that we are working with two theories with matched lattice spacings, $a = a'$, but mismatched fermion masses $m \neq m'$.  We first replace the two masses with the geometric mean $\tilde{m}$ and square root of the mass ratio $\rho$,
\beq
\tilde{m} \equiv \sqrt{m' m},\ \rho \equiv \sqrt{\frac{m'}{m}},
\eeq
so that $m = \tilde{m}/\rho$, $m' = \tilde{m} \rho$.  Given a quantity $X$ with finite intercept in chiral perturbation theory, we can rewrite its expansion to NLO in the two theories,
\begin{align}
X'_m &= X' \left\{ 1 + z' \tilde{m} \rho \left[ \alpha'_{10,X} + \alpha'_{11,X} \log (z' \tilde{m} \rho)\right] \right\}, \\
X_m &= X \left\{ 1 + \frac{z \tilde{m}}{\rho} \left[ \alpha_{10,X} + \alpha_{11,X} \log\left(\frac{z \tilde{m}}{\rho}\right)\right] \right\}.
\end{align}
Taking the ratio and discarding terms of higher order in $\tilde{m}$ and $\rho$, we have
\begin{align}
\frac{X'_m}{X_m} &= \frac{X'}{X} \left[ 1 + \tilde{m} \left( \beta_{10} + \beta_{11} \log \tilde{m}\right) \right], \\
\beta_{10} &= z' \rho (\alpha'_{10,X} + \alpha'_{11,X} \log(z' \rho) - \frac{z}{\rho} (\alpha_{10,X} + \alpha_{11,X} \log(z/\rho) ),\\
\beta_{11} &= z' \rho \alpha'_{11,X} - \frac{z}{\rho} \alpha_{11,X}.
\end{align}
This makes it clear that in order to recover the correct ratio of intercepts in the chiral limit, we should extrapolate to $\tilde{m} = 0$ while holding the mass ratio $\rho$ fixed.

\bibliographystyle{ws-procs975x65}
\bibliography{proc-neil-scgt12}

\end{document}